\documentclass[preprint,showpacs,preprintnumbers,english]{revtex4}
\usepackage{amsmath}
\usepackage{graphicx}
\usepackage{dcolumn}
\usepackage{bm}

\begin{document}

\title{Magnon mediated electric current drag across a ferromagnetic
insulator layer}
\author{Steven S.-L Zhang and Shufeng Zhang}
\affiliation{Department of Physics, University of Arizona, Tucson, AZ 85721}

\begin{abstract}

In semiconductor heterostructure, the Coulomb interaction is
responsible for the electric current drag between two 2-d electron
gases across an electron impenetrable insulator. For two metallic
layers separated by a ferromagnetic insulator (FI) layer, the
electric current drag can be mediated by a nonequilibrium magnon
current of the FI. We determine the drag current by using the
semiclassical Boltzmann approach with proper boundary conditions of
electrons and magnons at the metal-FI interface.
\end{abstract}
\date{\today}
\maketitle

The conventional Coulomb drag
effect\cite{Pogrebinskii77,Price88,Gramila91} occurs in
two-dimensional electron gases separated by an insulator barrier.
When one of the electron gas carries a current, the momentum
transfer due to Coulomb interaction leads to a small current in the
other electron gas. Recently, this current drag phenomenon has been
discovered in a different system with entirely different physical
mechanisms \cite{Kajiwara10}: when an electric current is injected
into a Pt bar deposited on a magnetic insulator Yttrium-Iron-Garnet
(YIG) film, it is found that a small electric voltage is induced in
the other Pt bar, which is also deposited on the same YIG film but
is located several millimeters away from the current carrying Pt
bar. The authors \cite{Kajiwara10} attributed their finding to the
combined effects of spin transfer torque (STT)
\cite{Slonczewski,Berger96} and spin pumping
\cite{Tserkovnyak02,Heinrich11}: the spin Hall \cite{Hirsch99}
current generated by the electric current in one Pt layer (Pt is a
known material with a large spin Hall angle) is absorbed by the FI
and for a sufficiently large STT, the magnetic moment of the FI
begins precessing. The precessing FI pumps out a spin current to the
other Pt layer. Finally, the resulting spin current converts to an
electron current due to the inverse spin Hall effect
\cite{Saitoh06,Valenzuela06,Kimura07,Madami11,Wang11,Xiao12}.

In this Letter, we propose a different geometry in which the magnon
current flows normal to the plane of the layers throughout the
structure. We show that the electron spin current in the metallic
layers induces a nonequilibrium magnon current in the FI layer. By
using semiclassical Boltzmann approach for electrons and magnons, we
are able to self-consistently determine these currents and thereby
obtain the drag current for given geometrical and material
parameters. The resulting drag current is several orders of
magnitude larger than that in the nonlocal geometry in Ref.
\cite{Kajiwara10}

\begin{figure}[tbp]
\includegraphics[width=8cm]{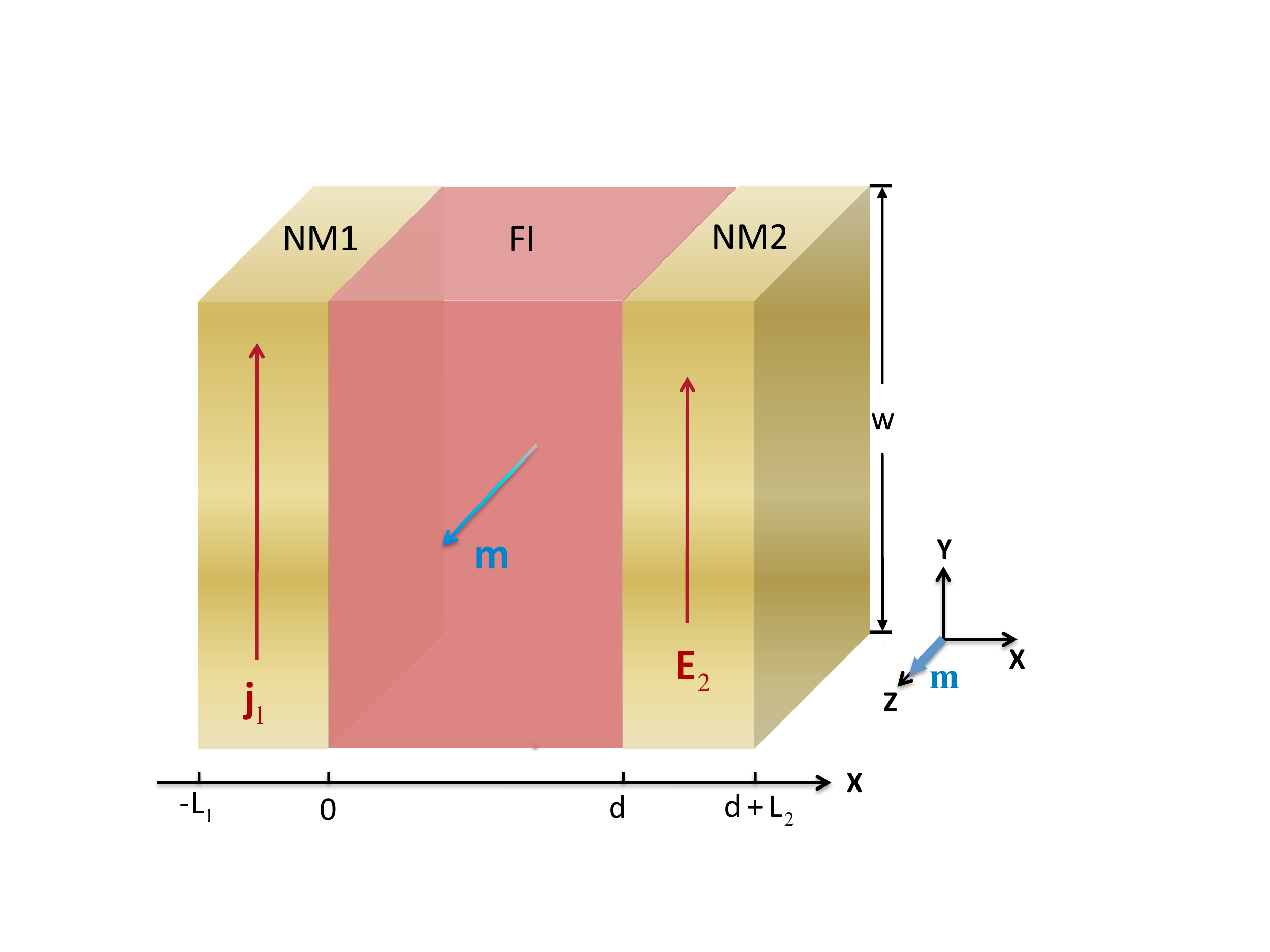}
\caption{Schematics of the NM/FI/NM trilayer structure.}
\end{figure}

To be more specific, we consider a simple trilayer structure, shown
in Fig. 1 schematically, where a ferromagnetic insulating (FI) layer
is sandwiched by two heavy metal films (NM1 and NM2) such as Pt and
Ta. A charge current parallel to the plane of the layers is injected
in the layer NM1. To determine the drag current in the layer NM2, we
first establish transport equations for each layers and then find
proper boundary conditions to solve the transport coefficients.

{\em Electron current and spin accumulation in Metallic layers}. For
the NM layers, a spin dependent Ohm's law has been well established
and may be written in the following form \cite{szhang00},
\begin{equation}
\mathbf{\hat{j}}=\frac{c}{2}\mathbf{\hat{E}+}\frac{c_{h}}{4}\left(
{\bf \hat{E}}\times \mbox{\boldmath $\sigma -\sigma$}\times
\mathbf{\hat{E}}\right)
\end{equation}%
where the spinor current density $\hat{\bf j}$ and the electric
field $\hat{\bf E}$ are $2\times 2$ vector matrices in spin space,
$\mbox{\boldmath $\sigma$}$ is a Paul vector matix, and $c$ and
$c_{h}$ are the electric conductivity and spin Hall conductivity
respectively. The second term is the spin Hall current whose
anti-symmetric form is essential for $\mathbf{\hat{j}}$ to be an
Hermitian in spin space (also noted that ${\bf \hat{E}}\times
\mbox{\boldmath $\sigma $}\neq - \mbox{\boldmath $\sigma$}\times
\mathbf{\hat{E}}$ due to non-communitivity of the Pauli matrices).
The electrical field is related to the spinor chemical potential
$\hat{\mu}$ via $\hat{\bf E}=-(1/e) \mbox{\boldmath ${\nabla }$}
\hat{\mu}$ where $e(<0)$ denotes the electron charge. While it is
possible to work with an arbitrary choice of the spin quantization,
we proceed below to a special case where the magnetic moment of the
FI is oriented in the $z$-direction and the electric current flows
in the $y$ direction. If we choose the spin quantization axis
parallel to the $z$-axis, one can simply work on the two-component
(spin-up and spin down) form of the Ohm's law; i.e.,
\begin{equation}
j_y^{\alpha} (x) =\frac{c}{2} E_y^{\alpha} (x) -\alpha \frac{c_h}{2}
E_x^{\alpha} (x)
\end{equation}
and \begin{equation} j_x^{\alpha} (x) = \frac{c}{2} E_x^{\alpha} (x)
+ \alpha \frac{c_h}{2} E_y^{\alpha} (x)
\end{equation}
where $\alpha = \pm 1 $ represent spin up an down. To determine the
spin dependent electric field, we recall the spin diffusion equation
\cite{Valet}
\begin{equation}
\frac{d^{2}}{dx^{2}}[ \mu^{\uparrow} (x)-\mu^{\downarrow} (x)]
=\frac{\mu^{\uparrow} (x)-\mu^{\downarrow} (x)}{\lambda_{sf}^{2}}
\end{equation}%
and its solution
\begin{equation}
 \mu^{\uparrow} (x)-\mu^{\downarrow} (x) =A_i
 e^{-x/\lambda_{sf}}+B_i
e^{x/\lambda_{sf}}
\end{equation}%
where $\lambda_{sf}$ is the electron spin diffusion length, and the
constants $A_i$ and $B_i$ ($i=1$ for the NM1 layer and $i=2$ for
NM2) are determined by the boundary conditions. Although these
equations apply to both NM1 and NM2 layers, they have different
constraints set by experimental measurement. For the NM1 layer, we
take $E_y^{\alpha} (x)=E_{ext}$ where $E_{ext}$ is the applied
electric field in the NM1 layer, while $\int_{NM2} dx j_y (x) =0$ in
the open circuit of the NM2 layer.

{\em Magnon current and magnon accumulation in the} FI {\em layer}.
For the FI layer, we start with a general magnon Boltzmann equation
in
the presence of spatially dependent temperature $T(x)$ and magnetic field $\mathbf{%
H}(x)$,
\begin{equation}
v_{x}\frac{\partial N_{m}}{\partial x}+v_{x}\frac{\partial N_{m}}{\partial T}%
\frac{dT}{dx}+v_{x}\frac{\partial N_{m}}{\partial \mathbf{H}}\cdot \frac{d%
\mathbf{H}}{dx}+\mathbf{\dot{q}\cdot }\frac{\partial N_{m}}{\partial \mathbf{%
q}}=-\left( \frac{\partial N_{m}}{\partial t}\right) _{scatt.}
\end{equation}%
where $N_{m}\left( x,\mathbf{q,}T(x),\mathbf{H}(x)\right) $ is the
magnon distribution. The first term describes magnon diffusion. The
second and third terms are responsible for the magnon transport in
the presence of temperature and magnetic field gradients, which have
been recently studied in the content of spincalorics
\cite{Uchida10,Bauer10,Loss03,Xing04}. The last term on the left
side of Eq.~(6) is associated with acceleration of magnons by
external forces such as a confining potential at boundary
\cite{Matsumoto11}. The scattering term on the right side of the
Eq.~(6) may be modeled by the relaxation time approximation
\begin{equation}
\left( \frac{\partial N_{m}}{\partial t}\right)
_{scatt.}=\frac{N_{m}-\bar{N}_{m}}{\tau
_{m}}+\frac{N_{m}-N_{m}^{0}}{\tau _{th}}
\end{equation}%
where $\bar{N}_{m}(x)=\int d\mathbf{q}N_{m}(x,\mathbf{q})/\int
d\mathbf{q}$ is the momentum averaged magnon distribution while $N_{m}^{0}(x,\mathbf{q})=%
\frac{1}{e^{\epsilon _{\mathbf{q}}/k_{B}T(x)}-1}$ is the local
equilibrium magnon distribution,
where $\epsilon _{\mathbf{q%
}}=D\mathbf{q}^{2}+\triangle _{g}$ is the magnon dispersion, $D $ is
the spin wave stiffness, $\triangle _{g}$ is the spin wave gap, and
$v_{x}=\frac{1}{\hbar }\frac{\partial \epsilon _{%
\mathbf{q}}}{\partial q_{x}}$ is the $x$ component of the magnon
velocity. The first relaxation term describes those processes which
conserve the number of magnons. For example, magnon scattering by a
paramagnetic impurity has the form of $V_{\bf qq'}a_{\bf q}^+a_{\bf
q'} $; i.e., the impurity or surface roughness
\cite{Sparks64,Mills03} scatters the magnon ${\bf q'}$ to the magnon
${\bf q}$. As long as we neglect the wave number dependence of the
scattering matrix $V_{\bf qq'}$, this process can be modeled by the
first term of Eq.~(7). The second term of Eq.~(7) does not conserve
the number of magnons. The magnon absorption and emission relax the
nonequilibrium magnons to equilibrium ones, e.g., magnon-phonon
interaction \cite{Suhl65}.

For the present system, we consider uniform temperature and magnetic
field, and there is no external force on magnons. Then, Eq.~(6) and
(7) reduce to
\begin{equation}
v_{x}\frac{\partial N_{m}\left( x,\mathbf{q}\right) }{\partial x}=-\frac{%
N_{m}\left( x,\mathbf{q}\right) -\bar{N}_{m}\left( x\right) }{\tau _{m}}-%
\frac{N_{m}\left( x,\mathbf{q}\right) -N_{m}^{0}\left( \mathbf{q}\right) }{%
\tau _{th}}.
\end{equation}
\bigskip
We may proceed to solve $N_m$ by the same way as for the electron
distribution in magnetic multilayers \cite{Valet}. Particularly, one
may expand the nonequilibrium distribution by the Legendre
polynomials,
\begin{equation}
N_{m}(x,\mathbf{q})=N_{m}^{0}(\mathbf{q})+\frac{\partial N_{m}^{0}(\mathbf{%
q})}{\partial \epsilon _{\mathbf{q}}}\left[ \mu_m (x) +
\underset{n=1}{\overset{\infty }{\sum }}g^{(n)}(x)P_{n} \left( \cos
\theta \right) \right]
\end{equation}%
where $\mu_m (x)$ is the $n=0$ component of the nonequilibrium
distribution and $\theta $ is the angle between $\mathbf{q}$ and $x$
axis. By placing the above equation into Eq.~(8) and by utilizing
the orthogonality property of the Legendre polynomials, one can
arrive at a series of algebraic equations for the coefficients
$g^{(n)} (x)$. In the supplemental material, we show the solutions
in some limiting cases. Once the distribution functions $N_m ({x,\bf
q})$ are determined, we can find the magnon accumulation and magnon
current via
\begin{equation}
j_{m} (x)=\frac{-2\mu_B}{\left( 2\pi \right) ^{3}}\int d\mathbf{q}
v_x N_m ({\bf q},x); \; \; \; \delta n_m (x) =\frac{1}{\left( 2\pi
\right) ^{3}}\int d\mathbf{q} \left[ N_m ({\bf q},x)-N_m^0 ({\bf
q})\right]
\end{equation}%
where $\mu_B$ is the Bohr magneton. Note that a magnon carries spin
moment $-\gamma \hbar(=-2\mu_B)$ where $\gamma$ is the gyromagnetic
ratio.

We may further simplify the solution of the non-equilibrium magnon
distribution by discarding high orders ($n \ge 2$) of the
polynomials. Consequently, we find a local relation between magnon
accumulation and magnon current,
\begin{equation}
\frac{d}{dx}j_{m}(x)=2\mu_B \frac{\delta n_{m}(x)}{\tau _{th}}
\end{equation}%
and%
\begin{equation}
j_{m}(x)=\frac{2\mu_B \tau_{m}}{3}
\frac{I_{2}}{I_{0}}\frac{d}{dx}\delta n_{m}(x)
\end{equation}%
where $I_{n}$ are integration
constants $I_{n}\equiv \frac{1}{(2\pi)^3}\int d^{3}qv^{n}\frac{\partial N_{m}^{0}(\mathbf{q})}{%
\partial \epsilon _{\mathbf{q}}}$. We point out that this local current expression is
valid in the limit $\tau _{th}\gg \tau _{m}$ which is a good
approximation for ferromagnets \cite{Suhl65} (see supplemental
material). By combining Eqs. (11) and (12), we obtain the
diffusion equation for nonequilibrium magnons,%
\begin{equation}
\frac{d^{2}}{dx^{2}}\delta n_{m}(x)-\frac{\delta n_{m}(x)}{l_{m}^{2}}=0
\end{equation}%
where the magnon diffusion length is defined as $l_{m}=\sqrt{\frac{I_{2}}{3I_{0}}%
\tau _{th}\tau _{m}}$. At room temperature ($T=300K$), for YIG with $%
\triangle _{g}\sim 10^{-5}eV$, $\tau _{m}\sim 10^{-7}s$ and $\tau
_{th}\sim 10^{-6}s$, $l_{m}$ is estimated at $0.05 cm$, consistent
with the
measurement \cite{Schneider08}. Equation (13) has the general solution,%
\begin{equation}
\delta n_{m}(x)=A_{F}e^{-x/l_{m}}+B_{F}e^{x/l_{m}}
\end{equation}%
and thus the magnon current density reads
\begin{equation}
j_{m}(x)=\frac{2\mu_B l_{m}}{\tau _{th}}\left(
-A_{F}e^{-x/l_{m}}+B_{F}e^{x/l_{m}}\right)
\end{equation}

{\em Boundary conditions}. The outer-boundary conditions at $x=-L_1$
and $x= d+L_2$, where $L_1$, $d$ and $L_2$ represent the thicknesses
of the layers of NM1, FI and NM2, are $j_x^{\uparrow,\downarrow}
(-L_1)=j_x^{\uparrow,\downarrow} (d+L_2) =0$. The boundary
conditions at the metal-FI interfaces depend on the interaction
between electrons and magnons. Here we assume a s-d type interaction
$-J_{sd} \mbox{\boldmath $\sigma$}\cdot {\bf S}_i$ where $
\mbox{\boldmath $\sigma$}$ is the itinerant electron spin of the
metal layer and ${\bf S}_i$ is the local spin of the FI layer at the
interfaces. The interaction conserves total angular momentum and
thus the first boundary condition is the continuity of total spin
current at the interfaces; i.e.,
\begin{equation*}
(-\mu_B/e)\left[j_x^{\uparrow} (0^-) - j_x^{\downarrow} (0^-)\right]
=j_m (0^+)
\end{equation*}%
and%
\begin{equation}
j_m (d^-) =(-\mu_B/e)\left[j_x^{\uparrow} (d^+) - j_x^{\downarrow}
(d^+)\right]
\end{equation}%
The total angular momentum current conservation simply states that
electron spin current in the metals must be converted into magnon
current in the FI layer at the interfaces. If the interfaces have
magnetic roughness, the spin-flip scattering by magnetic impurities
can transfer spin angular momentum to lattice via spin-orbit
coupling. In this case, the outgoing spin current would be reduced
\cite{footnote}.

The other boundary conditions at the interfaces should relate the
electron spin accumulation to the non-equilibrium magnon density.
Within the s-d model, one can treat the electron spin density as an
effective magnetic field on the interface spin of the FI layer;
i.e., $H_{eff} = J_{sd} \delta {\bf m}_z$ and thus
\begin{equation*}
\mu^{\uparrow} (0^-) -\mu^{\downarrow} (0^-)=\varepsilon \delta
n_{m} (0^+)
\end{equation*}%
and
\begin{equation}
\varepsilon \delta n_{m} (d^-)=\mu^{\uparrow} (d^+)
-\mu^{\downarrow} (d^+)
\end{equation}%
where
\begin{equation}
\varepsilon =\frac{4\left( \pi D\right) ^{3/2}}{J_{sd}D\left(
\varepsilon _{F}\right) a_{0}^{3}\cdot
\sqrt{k_{B}T}Li_{\frac{1}{2}}\left( e^{-\Delta _{g}/k_{B}T}\right) }
\end{equation}%
$D\left( \varepsilon_{F}\right)$ is the electron density of state at
Fermi level, $a_{0}$ is the lattice constant of the NM layer and
$Li_{s}\left( z\right) \equiv \sum\nolimits_{k=1}^{\infty
}\frac{z^{k}}{k^{s}}$ is the polylogarithm. The detailed derivation
of $ \varepsilon$ is arranged in the supplemental material. We note
that Takahashi et al. have also proposed a boundary condition at the
interface \cite{Takahashi10} which relates magnon spin current to
spin accumulation; i.e., $ j_{m}\propto \mu^{\uparrow}
-\mu^{\downarrow}$. However, such boundary condition is unable to
self-consistently determine the magnon current. On the contrary, the
boundary conditions we have derived are able to uniquely determine
the spin and magnon currents throughout the structure. A rough order
of magnitude estimation of $\varepsilon$ can be readily obtained by
using the following plausible parameters appropriate for Pt/YIG/Pt
structure: $D\left( \varepsilon_{F}\right)\sim 3 n_e/2 \epsilon _F$,
$n_{e}=5\times 10^{22}$ $cm^{-3}$, $J_{sd}=1$ $eV$, $\epsilon _F=5$
$eV$, $a_{0}=4$ ${\AA}$, and $D=6$ $meV\cdot nm^{2}$, and thus
$\varepsilon \sim 0.2$ $meV\cdot nm^{3}$.

With the above boundary conditions, the constants in Eq.~(5) and
(14) can be readily determined. If one uses an Ampere meter
\cite{voltmeter} to measure the average (measured) induced electric
current density $j_y^{(2)} = (1/L_2)\int dx j_y^{(2)} (x)$ in NM2
layer, we find the ratio between the induced current and the
injected current magnitude $\eta \left(\equiv
j_y^{(2)}/j_y^{(1)}\right)$ is,
\begin{equation}
\eta =\left( \frac{c_h}{c} \right)^2 \left(
\frac{\lambda_{sf}}{L_2}\right) \frac{\operatorname{sech}\left(
\frac{L_2}{\lambda_{sf}}\right) \left[ \cosh \left( \frac{L_2}{
\lambda_{sf}}\right)-1 \right] }{\left[ b^{-1} + b \tanh \left(
\frac{L_{2}}{\lambda_{sf}}\right) \right] \sinh \left(
\frac{d}{l_{m}}\right) +\left[ 1+\tanh \left( \frac{L_{2}}{
\lambda_{sf}}\right) \right] \cosh \left( \frac{d}{l_{m}}\right) }
\end{equation}%
where $b=c\varepsilon \tau_{th}/(4e^{2}l_{m}\lambda_{sf})$, and $L_1
\gg \lambda_{sf} $ is assumed for simplicity. The first prefactor
$(c_h/c)^2 $ origins from the two successive conversions between
electric current and spin Hall current in NM1 and NM2 due to the
spin Hall and inverse spin Hall effect respectively. The second
prefactor $\lambda_{sf}/L_2$ indicates that the maximum range of the
current density in NM2 is $\lambda_{sf}$; i.e., if the thickness of
NM2 exceeds $\lambda_{sf}$, the average current density $j_y^{(2)}$
would be inversely proportional to $L_2$. Interestingly, when $L_2$
is much smaller than $\lambda_{sf}$, the induced electric current is
also small; this is because the spin current at the surface
$x=d+L_2$ is set to zero and thus the self-consistent calculation
demands a small current throughout the NM2 layer. In Fig.2, we show
$\eta$ as a function of the thickness of the metal layer (NM2) for
Pt/YIG/Pt trilayers with several different YIG layer thicknesses. We
choose the material parameters as follows: Pt layer conductivity
$c_{Pt}\sim 0.1\left( \mu \Omega \cdot cm\right) ^{-1}$, spin
diffusion length $\lambda_{sf}=7$ $nm$ and the spin Hall angle
$c_h/c=0.05$ \cite{Buhrman11}; magnon diffusion length $l_{m}=0.05$
$cm$ and magnon relaxation time $\tau _{th}=10^{-6}$ $s$. We see
that $\eta $ decreases as the thickness of the YIG layer increases
due to the decay of magnon diffusion current. Also for fixed YIG
layer thickness, $ \eta $ reaches its maximum around
$L_2=\lambda_{sf}$. The peak value of $\eta $ is of the order of
$10^{-4}$. If the injected current density is $10^{6}$ $A/cm^{2}$,
the induced voltage of a Pt bar with its length $w=1$ cm would be
$V^{(2)}=wj_y^{(2)}/c\sim 1 $ mV. If one replaces Pt with Ta which
has larger spin Hall angle of $0.1$ \cite{Buhrman12}, $V^{(2)}$ can
be further increased by a factor of 4, which is rather significant
and easily detectable experimentally.
\begin{figure}[tbp]
\includegraphics[width=8cm]{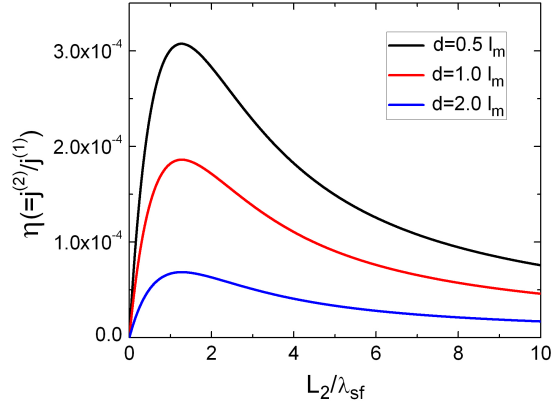}
\caption{The ratio of the average induced current density and the
injected current density as a function of the NM2(Pt) layer
thickness for three different thicknesses of the FI (YIG) layer.}
\end{figure}

Finally we comment on the relation of our calculation with the
experimental measurement \cite{Kajiwara10}. In their experiments,
when the first Pt layer injects a spin current to the FI layer, the
magnons propagate {\em in the plane of the FI layer} in order to
reach the second Pt. While there is a similar non-equilibrium magnon
density buildup near the second Pt layer, the {\em direction} of the
magnon current and the {\em gradient} of the magnon density are in
the plane of the layer. In another word, there is neither magnon
current nor magnon density gradient in the direction perpendicular
to the layer such that the second Pt layer is unable to receive any
spin angular momentum from the FI layer. Thus, we conclude that the
nonlocal setup in the experiment \cite{Kajiwara10} is not relevant
to our theory. In the conventional nonlocal metallic spin valve,
however, one does observe a voltage change of the entire detection
bar due to the spin accumulation (not the spin current or gradient
of the spin accumulation) in the channel. In the present case, we
derive the induced current in the second Pt bar which is related to
the spin current (or magnon density gradient) in the direction
perpendicular to the layer. Furthermore, the observed current in the
experiment\cite{Kajiwara10} has been attributed to the STT and spin
pumping, which is several orders of magnitude smaller than what we
predict in our geometry.

This work is supported by NSF.

\newpage
\section*{SUPPLEMENTAL MATERIALS}

\subsection{Magnon diffusion equation in ferromagnetic insulator}

\bigskip

In this section, we derive the magnon diffusion equation, Eq.~(13),
from the magnon Boltzmann equation, Eq.~(8). By placing Eq.~(9) into
Eq.~(8), we have
\begin{align}
& v_{x}\frac{\partial N_{m}^{0}(\mathbf{q})}{\partial \epsilon _{\mathbf{q}}}%
\frac{\partial g(x,\mathbf{q})}{\partial x}+\frac{\partial N_{m}^{0}(\mathbf{%
q})}{\partial \epsilon _{\mathbf{q}}}\left( \frac{1}{\tau _{m}}+\frac{1}{%
\tau _{th}}\right) g(x,\mathbf{q})  \notag \\
& =-v_{x}\frac{\partial N_{m}^{0}(\mathbf{q})}{\partial \epsilon _{\mathbf{q}%
}}\frac{d\mu_m \left( x\right) }{dx}-\frac{\partial N_{m}^{0}(\mathbf{q})}{%
\partial \epsilon _{\mathbf{q}}}\frac{\mu_m \left( x\right) }{\tau _{th}}-%
\frac{\mu_m \left( x\right) }{\tau _{m}}\left[ \frac{\partial N_{m}^{0}(%
\mathbf{q})}{\partial \epsilon _{\mathbf{q}}}-\bar{\frac{\partial N_{m}^{0}}{%
\partial \epsilon _{\mathbf{q}}}}\right] -\frac{N_{m}^{0}(\mathbf{q})-\bar{N}%
_{m}^{0}}{\tau _{m}}  \tag{A1}
\end{align}%
\newline
where $\bar{\frac{\partial N_{m}^{0}}{\partial \epsilon _{\mathbf{q}}}}%
\equiv \int d\mathbf{q}\frac{\partial
N_{m}^{0}(\mathbf{q})}{\partial
\epsilon _{\mathbf{q}}}/\int d\mathbf{q}$, $\bar{N}_{m}^{0}\equiv \int d%
\mathbf{q}N_{m}^{0}(\mathbf{q})/\int d\mathbf{q}$, and $g(x,\mathbf{q})=\underset{n=1}{\overset{\infty }{\sum }}g^{(n)}(x)P_{n}%
\left( \cos \theta \right)$ is the Legendre polynomial expansion of
the nonequilibrium magnon distribution when a rotational symmetry is
assumed.

The non-equilibrium magnon density and magnon current, as defined in
Eq.~(10), are simply the zeroth and first components of the Legendre
polynomials,
\begin{equation}
\delta n_{m}(x) =4 \pi I_0 \mu_m (x)  \tag{A2}
\end{equation}%
and
\begin{equation}
j_{m}(x)=-\frac{8\pi \mu_B }{3}I_{1}g^{(1)}(x) \tag{A3}
\end{equation}%
where we have defined
\begin{equation}
I_{n}\equiv \frac{1}{(2\pi)^3} \int d^{3}qv^{n}\frac{\partial N_{m}^{0}(\mathbf{q})}{%
\partial \epsilon _{\mathbf{q}}} \tag{A4}
\end{equation}
$v=({\bf q}/q)\cdot \mbox{\boldmath $\nabla$} \epsilon_m ({\bf
q})/\hbar$ is the magnitude of magnon velocity.

The relations among $g^{(n)}$ can be readily obtained by multiplying
Eq.~(A1) by $v_x^n$ and integrating over ${\bf q}$,
\begin{equation}
\frac{d}{dx}j_{m}(x)=2 \mu_B \frac{\delta n_{m}(x)}{\tau _{th}},
\text{ \ \ \ }(n=0) \tag{A5}
\end{equation}
\begin{equation}
\frac{2I_{2}}{5}\frac{d}{dx}g^{(2)}(x)+I_{1}\left( \frac{1}{\tau _{m}}+\frac{%
1}{\tau _{th}}\right) g^{(1)}(x)=-I_{2}\frac{d}{dx}\mu_m \left( x\right) ,%
\text{ \ \ \ }(n=1)  \tag{A6}
\end{equation}%
and
\begin{equation}
I_{n+1}\left[ \frac{n}{2n-1}\frac{
dg^{(n-1)}}{dx}+\frac{n+1}{2n+3}\frac{ dg^{(n+1)}}{dx}\right]
+I_{n}\left( \frac{1}{\tau _{m}}+ \frac{1}{\tau _{th}}\right)
g^{(n)}=0,\text{ \ \ \ }(n \ge 2) \tag{A7}
\end{equation}%
where we have used the following orthogonal relations for the Legendre polynomials%
\begin{equation}
\overset{1}{\underset{-1}{\int }}duP_{n^{\prime }}\left( u\right)
P_{n}\left( u\right) =\frac{2}{2n+1}\delta _{n,n^{\prime }} \tag{A8}
\end{equation}%
and%
\begin{equation}
\overset{1}{\underset{-1}{\int }}duP_{1}\left( u\right) P_{n^{\prime
}}\left( u\right) P_{n}\left( u\right) =\frac{2\left( n+1\right)
}{\left( 2n+1\right) \left( 2n+3\right) }\delta _{n^{\prime
},n+1}+\frac{2n}{\left( 2n+1\right) \left( 2n-1\right) }\delta
_{n^{\prime },n-1}  \tag{A9}
\end{equation}%
At this point, the problem of solving the magnon distribution is
equivalent to determining the infinite function series $g^{(n)}(x)$.
To simplify the problem, we assume $\tau_{th} \gg \tau_m$, which is
generally valid for most insulating ferromagnetic materials
\cite{Suhl65}. In this case, the ratio of $g^{(n+1)}/g^{(n)}$ is
about $\sqrt{\tau_m/\tau_{th}}$. To see it, we first neglect
$g^{(2)}$ in Eq.~(A6) and substitute the resulting expression for
$g^{(1)}$ into Eq.~(A5). One immediately sees that the spatial
derivatives of both $\mu_m$ and $j_m$ scale as $1/\sqrt{\tau_m
\tau_{th}}$. Putting this scaling back to Eq.~(A6) and (A7), we
confirm that neglecting $g^{(n)}$ ($n\ge 2$) is justified as long as
$\tau_{th} \gg \tau_m$. Combining Eqs.~(A2), (A3), (A5) and (A6), we
arrive at the magnon diffusion equation in the main text, Eq.~(13).

We emphasize that the validity of the magnon diffusion equation
rests on the condition that the magnon conserving scattering is much
stronger than the magnon non-conserving scattering, i.e., $\tau_{th}
\gg \tau_m$; this condition is similar to the spin diffusion
equation for electrons where the momentum scattering
(spin-conserving) is stronger than the spin-flip scattering
\cite{Valet93}. Thus, the magnon diffusion equation can be used for
the length scale larger than the spin-conserving scattering length
(e.g., it is the mean free path in the electron case) which is
considered much smaller than the magnon (spin) diffusion length.

\subsection{\protect\bigskip Boundary conditions at metal-FI
interfaces}

\bigskip

In this section, we derive the boundary conditions Eqs.~(16) and
(17) by using the $s-d$ exchange coupling
\begin{equation}
\hat{H}=-J_{sd}\underset{i}{\mathbf{\sum }}\int d^{3}r
\mathbf{s}\left( \mathbf{r}\right) \cdot \mathbf{S}\left(
\mathbf{R}_{i}\right) \delta \left( \mathbf{r-R}_{i}\right),
\tag{B1}
\end{equation} where ${\bf S}(\textbf{R}_i)$ are the localized spins of FI at the interface
in contact with the conduction electron spins
$\textbf{s}(\textbf{r})$ of the metal. The above s-d interaction
contains a spin-flip scattering processes of $J_{sd} (c_{{\bf k}
\downarrow}^+ c_{{\bf k-q}\uparrow} a_{\bf q} +h.c.)$ and a
non-spin-flip process $J_{sd}c_{{\bf k+q'} \sigma}^+ c_{{\bf k+q}
\sigma} a_{\bf q}^+ a_{\bf q'}$\cite{White68}. For the spin-flip
process, the electron spin loss (gain) leads to the magnon
generation (annihilation), but the total angular momentum is
conserved; this leads to the first interface boundary condition of
Eq.~(16). For the non-spin-flip processes, we use the simple mean
field approach, i.e., the conduction electron spin density is
treated as an effective magnetic field acting on the interficial
spins of the FI layer; i.e.,
\begin{equation}
\delta H_{eff} =-\frac{J_{sd}D\left( \varepsilon _{F}\right)
v_{0}}{g\mu _{B}}\left(\mu^{\uparrow} -\mu^{\downarrow}\right)
\tag{B2}
\end{equation}
where $D(\varepsilon _{F})$ is the electronic density of states and
$v_0$ is the volume of primitive cell of the metal layer. The
consequence of this effective field is to induce, in linear
response, a deviation in longitudinal magnetization%
\begin{equation}
\delta M_{z} =\chi_{zz}\delta H_{eff} \tag{B3}
\end{equation}%
where $\chi_{zz}$ is the longitudinal susceptibility for interface
spins of the FI layer and $\delta M_{z}$ is related to the magnon
density $\delta M_{z}(x)=-2\mu _{B}\delta n_{m}(x) $. Thus, the
relation between spin accumulation of non-equilibrium electron
density and nonequilibrium magnon density at the metal-FI interface
is
\begin{equation}
\mu^{\uparrow} -\mu^{\downarrow} =\varepsilon \delta n_{m} \tag{B4}
\end{equation}%
where
\begin{equation}
\varepsilon =\frac{g\mu _{B}^{2}}{J_{sd}D\left( \varepsilon
_{F}\right) \chi _{zz}v_{0}} . \tag{B5}
\end{equation}%
$\chi _{zz}$ can be derived in linear response by $\chi _{zz}=\left. \frac{%
\partial M_{z}}{\partial H}\right\vert _{H\rightarrow 0}$ with $M_{z}=M_{S}-%
\frac{2\mu _{B}}{\left( 2\pi \right) ^{3}}\int
d\mathbf{q}N_{m}^{0}\left( \mathbf{q}\right) $. In the limit
$Dq_{\max }^{2}\gg k_{B}T$, we find
\begin{equation}
\chi _{zz}=\frac{g\mu _{B}^{2}\sqrt{k_{B}T}}{4\left( \pi D\right) ^{3/2}}Li_{%
\frac{1}{2}}\left( e^{-\Delta _{g}/k_{B}T}\right)  \tag{B6}
\end{equation}%
where $Li_{s}\left( z\right) \equiv \sum\nolimits_{k=1}^{\infty }\frac{z^{k}%
}{k^{s}}$ is the polylogarithm. Inserting Eq.(B6) into (B5), we get
the final result $\varepsilon $ of Eq.~(18).


\begin{thebibliography}{99}
\bibitem{footnote}
The spin current loss at the interfaces depends on the relative
strenths of electron-impurity scattering and electron-magnon
scattering. We may introduce a phenomenological spin-loss
coefficient $\beta$ ($0 \le \beta \le 1$) by timing $\beta$ to the
left sides of Eq.~(16). Our final result, Eq.~(19), would be reduced
by a factor of $\beta^2$.
\bibitem{voltmeter}
If we use a voltmeter, the induced electric field is related to the
induced current by $E_y (x)= c^{-1}j_y (x)$.
\end{thebibliography}
\end{document}